\documentclass[11pt,letterpaper]{article}
\usepackage{color}
\usepackage{amsmath}
\usepackage{amsthm}
\usepackage{amsfonts}
\usepackage{amssymb}
\usepackage{graphicx}
\usepackage{float}
\usepackage{geometry}
\usepackage[utf8]{inputenc}

\usepackage[all]{xy}
\usepackage{amscd}
\usepackage{amsthm}
\usepackage{longtable}
\usepackage{mathrsfs}
\usepackage{amsmath}
\usepackage{amssymb}
\usepackage{enumitem}
\usepackage[table,dvipsnames]{xcolor}

\usepackage{soul}

\usepackage{booktabs}

\usepackage{array}

\usepackage{pdflscape}
\usepackage{listings}
\lstdefinestyle{myCustomRStyle}{
  language=R,
  numbersep=10pt,
  basicstyle=\ttfamily\tiny,
  tabsize=4,
  showspaces=false,
  showstringspaces=false,
  backgroundcolor = \color{white},
  rulecolor=\color{white},
  breakatwhitespace=false,
  breaklines=true,
  keepspaces=true,
  showspaces=false,
  showstringspaces=false,
  showtabs=false
}
\usepackage{xcolor}
\usepackage{colortbl}

\definecolor{Gray}{gray}{0.80}

\usepackage{color}

\newcommand{\cf}[1]{\textcolor{black}{#1}}
\newcommand{\ap}[1]{\textcolor{black}{#1}}

\definecolor{dkgreen}{rgb}{0,0.6,0}
\definecolor{gray}{rgb}{0.5,0.5,0.5}
\definecolor{mauve}{rgb}{0.58,0,0.82}

\usepackage{caption}

\usepackage[utf8]{inputenc}
\usepackage[english]{babel}

\usepackage[thinlines]{easytable}

\usepackage{tikz}
\usetikzlibrary{positioning}

\usepackage{pdfpages}

\usepackage{listings} 
\usepackage{xcolor}

\usepackage{comment}
\usepackage{color}
\usepackage[backref=page]{hyperref}
\hypersetup{
 colorlinks=true,
 linkcolor=blue,
 filecolor=magenta,
 citecolor=darkRed,
 urlcolor=cyan,
 pdftitle={
 Comparison of open-source software for producing directed acyclic graphs},
 }

\definecolor{codegreen}{rgb}{0,0.6,0}
\definecolor{codegray}{rgb}{0.5,0.5,0.5}
\definecolor{codepurple}{rgb}{0.58,0,0.82}
\definecolor{backcolour}{rgb}{0.95,0.95,0.92}
\definecolor{darkRed}{rgb}{0.60,.03,.03}

\lstdefinestyle{mystyle}{
    backgroundcolor=\color{backcolour},   
    commentstyle=\color{codegreen},
    keywordstyle=\color{magenta},
    numberstyle=\tiny\color{codegray},
    stringstyle=\color{codepurple},
    basicstyle=\ttfamily\footnotesize,
    breakatwhitespace=false,         
    breaklines=true,                 
    captionpos=b,                    
    keepspaces=true,                 
    numbersep=5pt,                  
    showspaces=false,                
    showstringspaces=false,
    showtabs=false,                  
    tabsize=2
}

\theoremstyle{definition}

\newtheorem*{counter example}{Counter Example}

\DeclareUnicodeCharacter{2008}{\textcolor{red}{BAD!!}}

\begin{document}

\def\spacingset#1{\renewcommand{\baselinestretch}{#1}\small\normalsize} \spacingset{1}

\title{ {\Large Comparison of open-source software for producing directed acyclic graphs}}
\author{ {\normalsize Amy J. Pitts, Charlotte R. Fowler }  }
\date{}

\clearpage \maketitle
\thispagestyle{empty}


\begin{center}
    Department of Biostatistics, Mailman School of Public Health, Columbia University 
\end{center}

\begin{abstract}

Many software \cf{packages} have been developed to assist researchers in drawing directed acyclic graphs (DAGs), \cf{each with unique functionality and usability}. We examine \ap{five} of the most common software to generate DAGs: Ti\textit{k}Z, DAGitty, ggdag, dagR, \ap{and igraph}. For each \cf{package,} we provide a general description of the \cf{its} background, analysis and visualization capabilities, and user-friendliness. \cf{Additionally in order to compare packages}, we produce two DAGs in each software, the first featuring a simple confounding structure, while the \cf{second} includes \cf{a} more complex structure with three confounders and a mediator. We provide recommendations for when to use each software depending on the user's needs. 

\end{abstract}	
\noindent%
 \vfill


\newpage 
\setcounter{page}{1}

\section{Introduction}
\cf{Research describing how to establish causal relationships has become of increased interest in many disciplines \cite{greenland1999dagdesc, miguel2023whatif, pearl2009causality, imbens2015causal, morgan2015counterfactuals, spirtes2000causation}, especially in cases where a randomized control experiment is not feasible.} One key tool to visualize hypothesized causal relationships, \cf{identify} where biases may arise, and \cf{inform} how to address them is a directed acyclic graph (DAG) \cite{greenland1999dagdesc, miguel2023whatif, morgan2015counterfactuals}. These graphs provide a display of the connections between the exposure, outcome, and other relevant variables. DAGs are employed across disciplines \cf{including} epidemiology \cite{suttorp2015graphical, shrier2008reducing, vanderweele2007four}, sociology \cite{knight2013causal, holland1988causal, kohler2023control},  education \cite{freedman1987others, costa2019mining, huynh2014applying}, and economics \cite{imbens2020potential, yang2014energy, awokuse2006export}. \ap{DAGs} consist of nodes and edges, where the nodes represent variables and the edges \cf{convey} direct causal effects by displaying an arrow leaving from the cause and pointing toward the effect. Importantly, a graph qualifies as \cf{a} DAG if no variable is an ancestor of itself, meaning no cycles occur in the graph, and each edge is pointed in \cf{a single} direction \cite{glymour2016causal}. For a DAG to be considered causal, \cf{it} is required to include all variables that are common causes of any two existing variables in the graph \cite{greenland1999dagdesc}.     

\cf{Developers have introduced software to produce DAGs across a variety of platforms.} DAGitty, dagR, ggdag, igraph, pcalg, and bnlearn are open\cf{-}source packages in R offering a range of plotting and analysis capabilities \cite{textor2016robust, barrett2018ggdag, csardi2006igraph, breitling2010dagr, kalischpcalg2012, bnlearn2017}. DAGitty offers both a browser-based platform and \cf{an} R package for creating, editing, and analyzing causal diagrams \cite{textor2016robust}. Ggdag extends the plotting functionality of DAGitty, and is tidyverse and ggplot compatible \cite{barrett2018ggdag}. DagR focuses on analysis and data simulation capabilities and provides a framework to draw, manipulate, and evaluate DAGs \cite{breitling2010dagr}. The R package igraph is designed for network analysis and is especially capable of handling large graph\ap{ical} systems \cite{csardi2006igraph}. Pcalg centers around causal structure learning and causal inference discovery but \cf{also has} some visualization features \cite{hauser2012Characterization, kalischpcalg2012}. The bnlearn package also focuses on causal discovery through \cf{B}ayesian network structure learning, as well as parameter learning \cite{bnlearn2017}. In Python\cf{,} three prominent libraries for causal graphs are causal-learn, causal discovery toolbox, and gCastle \cite{causallearn, kalainathan2019causal, zhang2021gcastle}. All three software \cf{packages} are focused on their algorithms for causal discovery \cf{but have} some limited DAG plotting capabilities. In the document preparation system \LaTeX, the graphing library Ti\textit{k}Z is commonly used to draw DAGs \cite{tantau2023tikz}. Quiver, a web\cf{-}based application, allows users to quickly draw a DAG through click \ap{and} drag motions and has \cf{the} functionality to export created DAGs to \LaTeX\ code \cite{quiver}. Causal fusion is \cf{a} similar web\cf{-}based application, however\cf{,} access to this resource requires an approved account \cite{causalfusion}. Tetrad, a free downloadable tool with over 30 year\cf{s of} history, creates, simulates, estimates, tests, and predicts causal and statistical models using DAGs \cite{ramsey2018tetrad}. This tool's functionality is very similar to pcalg and bnlearn. With this wide array of DAG drawing software, knowing which option is the most appropriate and easily implementable is a challenge. In 2006 Haughton et. al. provided a comparison paper \cite{haughton2006review}, that reviewed three statistical methods to illustrate DAGs (MIM, Tetrad, and WinMine). Since their publication statistical methods to illustrate DAGs have changed; indeed Tetrad is the only software compared by Haughton et. al. that remains maintained.

While numerous methods \cf{clearly} exist for designing, analyzing\cf{,} and visualizing a directed acyclic graph using software, there is no centralized resource comparing modern methods or providing recommendations. \cf{DAGs are a commonly used visual tool in publications}; indeed  Tennant et. al. analyzed a collection of 234 articles published between 1999-2017 that mentioned concepts related to DAGs and found that two-thirds of the articles made at least one DAG available \cite{tennant2021use}. However, Tennant et. al. noted that such DAGs ranged drastically in size, quality, and notation \cite{tennant2021use}. This is likely in part due to the lack of a comprehensive description of the available software for drawing DAGs. Here, we provide a \cf{guide that} highlights\cf{,} compares\cf{, and demonstrates how to employ each} DAG software. 

In our review\cf{,} we include open\cf{-}source software \cf{that} provide a manual of all features or published documentation and are able to be implemented directly in popular programming languages among statisticians, such as R or \LaTeX.  We additionally choose to restrict our review to packages focused on visualizing and producing DAGs, and thus exclude pcalg, bnlearn, causal-learn, causal discovery toolbox, and gCastle \cf{because} their main purpose is causal discovery and analysis. This restricts our scope to five resources: \ap{ Ti\textit{k}Z, DAGitty, ggdag, dagR, and igraph.} We evaluate the software across three categories: visual design, analysis capability, and utility. \cf{For visual design, we consider whether the software can include curved edges, subscripts, and math notation, the default visual settings,  design customization capabilities, and ability to allow for auto-generated and user\cf{-}specified node placements.} Next, to evaluate analysis capability we check for the presence of an exposure, outcome, and covariate framework; \cf{the }ability to identify ancestor/descendant relationships, conditional independencies, and minimally sufficient adjustment sets; and \cf{the} capacity to simulate data. Lastly, we evaluate the utility of the \ap{five} \cf{packages} by comparing the resources available, our experience of the learning curve, and \cf{the} required base software. In the methods section, we describe each software across evaluation criteria and compare performance between methods. In the discussion, we provide general recommendations \cf{specific to} the user's needs, and future directions for DAG software.

\section{DAG Producing Methods}

\begin{landscape}
    \begin{figure}
    \centering
    \begin{tabular}[t]{c |c |c |c |c } \toprule
         \textbf{Ti\textit{k}Z} &  \textbf{DAGitty} & \textbf{ggdag} & \textbf{dagR} & \textbf{igraph}\\

     \begin{tikzpicture}
\node[circle, draw] (C) at (0, 2) {C}; 
\node[circle, draw] (A) at (0, 0) {A};
\node[circle, draw] (Y) at (3, 0) {Y};
\draw[->, thick]
(C) edge (A)
(C) edge (Y)
(A) edge (Y);
\end{tikzpicture}  & 
\includegraphics[scale=0.3]{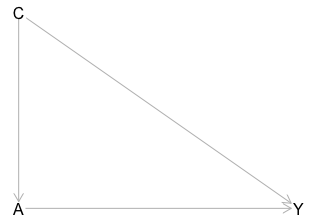} &

\includegraphics[scale=0.25]{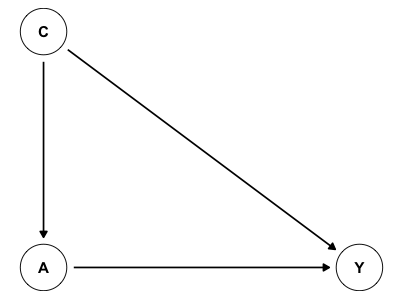} & \includegraphics[scale=0.29]{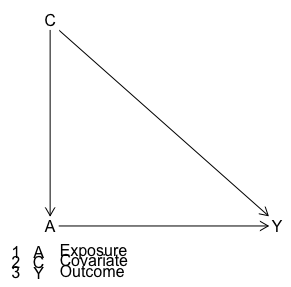} &
\includegraphics[scale=0.25]{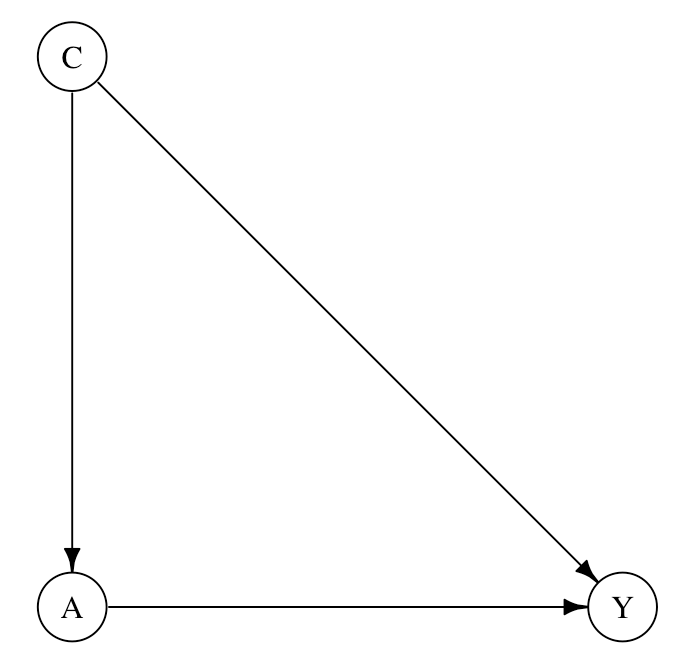}   \\ \hline
\begin{lstlisting}
\usepackage{tikz}
\tikzset{> = stealth}

\begin{document}
\begin{tikzpicture}
% nodes
\node[circle,draw] (C) at (0,2) {C}; 
\node[circle,draw] (A) at (0,0) {A};
\node[circle,draw] (Y) at (3,0) {Y};
% edges
\draw[->, thick]
(C)	edge (A)
(C)	edge (Y)
(A)	edge (Y);
\end{tikzpicture}
\end{document}
\end{lstlisting} & 
\begin{lstlisting}
library(dagitty)
dag1 <- dagitty( 
  "dag {
   A [exposure]
   Y [outcome]
   A -> Y
   C -> A 
   C -> Y}" )
coordinates(dag1) <-  list(
  x = c(C = 0, A = 0, Y = 1),
  y = c(C = 0, A = 1, Y = 1)
)
plot(dag1, lwd=2)
\end{lstlisting}
 & 
\begin{lstlisting}
library(ggdag)
dag1 <- dagify(
    A ~ C,
    Y ~ A + C,
    exposure = "A", 
    outcome = "Y",
    # location for each node
    coords = list(
      x = c(Y = 4, A = 2, C = 2),
      y = c(Y = 2, A = 2, C = 3)
    )
)
dag1 %>%
  ggplot(aes(x = x, y = y, 
    xend = xend, yend = yend)) +
  geom_dag_text(color = "black") +
  geom_dag_edges() +
  geom_dag_point(shape = 1) +
  theme_dag()
\end{lstlisting} & 
\begin{lstlisting}
library(dagR)
dag1 <- dag.init(
  covs = c(1), #1 per cov
  arcs = c(1,0, # C to A
           1,-1, # C to Y
           0,-1),# A to Y
  assocs = c(0),#directed arcs
  xgap = 0.04, 
  ygap = 0.05, 
  len = 0.1,
  x.name = "Exposure",
  cov.names = c("Covariate"),
  y.name = "Outcome",
  symbols = c("A", "C", "Y"),
  noxy = T
  )

#x, y-axis positions
dag1$x <- c(0, 0, 1) 
dag1$y <- c(0, 1, 0)
plot1 <- dag.draw(
  dag1, noxy = T, legend = T)
\end{lstlisting} & 
\begin{lstlisting}
library(igraph)
dag1 <- make_graph(
  edges = c("C","A", #C to A
            "C","Y", #C to Y
            "A","Y"),#A to Y 
  directed = TRUE)
plot(dag1, 
  layout =
    matrix(c(0, 0, 1,
             1, 0, 0), 
      nrow = 3, ncol = 2),
  vertex.size = 25,
  vertex.color = "white", 
  edge.color = "black", 
  vertex.label.color = "black", 
  edge.arrow.size = 0.6)
\end{lstlisting} \\ \bottomrule
    \end{tabular} 
    \caption{Simple confounding DAG example for each software. ``A'' denotes the exposure, ``Y'' the outcome, and ``C'' the counfounder. The code used to generate each DAG is included below the relative plot. 
    }
    \label{fig:dag1}
\end{figure}
\end{landscape}

\begin{landscape}
    \begin{figure}
    \centering
    \begin{tabular}[t]{c|c|c|c|c } \toprule 
         \textbf{Ti\textit{k}Z} &  \textbf{DAGitty} & \textbf{ggdag} & \textbf{dagR} & \textbf{igraph}\\
\begin{tikzpicture}[scale=.75]
\node[circle,draw] (C1) at (0, 2) {$C_1$}; 
\node[circle,draw] (C2) at (0, 0) {$C_2$}; 
\node[circle,draw] (C3) at (2, 2) {$C_3$}; 
\node[circle,draw] (A) at (0, 4) {$A$};
\node[circle,draw] (Y) at (4, 4) {$Y$};
\node[circle,draw] (M) at (2, 4) {$M$}; 
\draw[->, thick]
(C1) edge (A)
(C1) edge (M)
(C1) edge (Y)
(C1) edge (C2)
(C2) edge [bend left=40] (A)  
(C2) edge (M) 
(C2) edge [bend right=30] (Y)
(C3) edge (M) 
(C3) edge (Y) 
(A) edge [bend left=30] (Y)
(A) edge (M)
(M) edge (Y);
\end{tikzpicture}  & 
\includegraphics[scale=0.24]{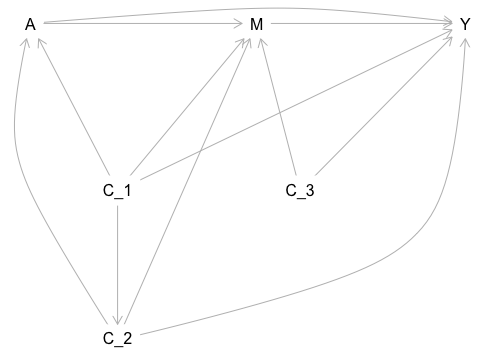} &
\includegraphics[scale=0.27]{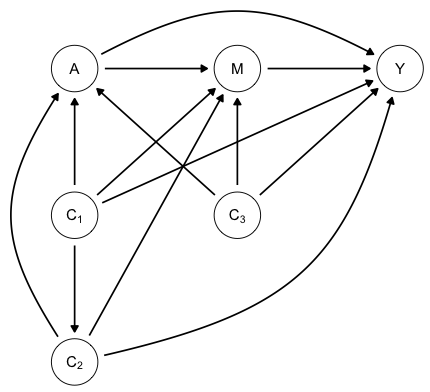} & 
\includegraphics[scale=0.32]{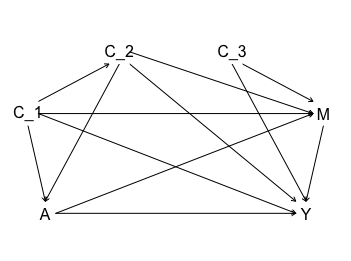} &
\includegraphics[scale=0.31]{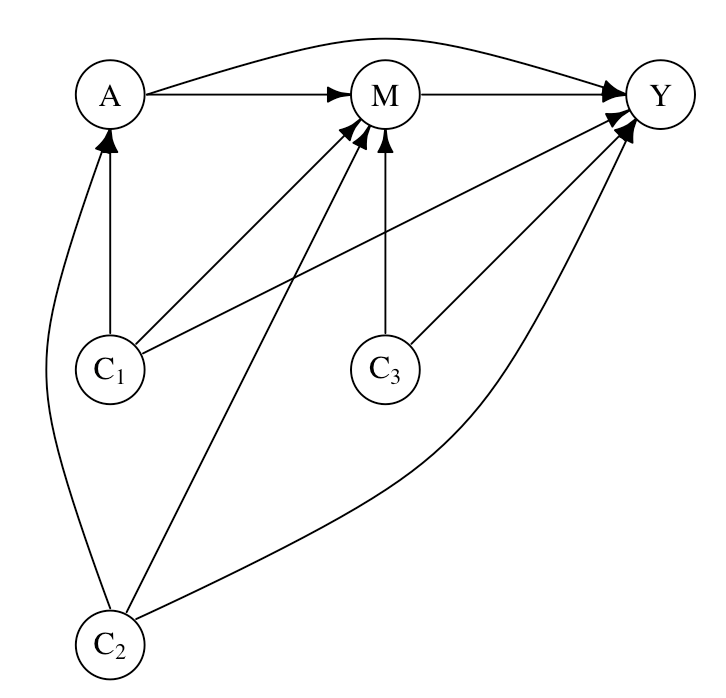}   \\ \hline 
\begin{lstlisting}
\usepackage{tikz}
\tikzset{> = stealth}

\begin{document}
\begin{tikzpicture}
% nodes
\node[circle,draw](C1) at (0,2){$C_1$}; 
\node[circle,draw](C2) at (0,0){$C_2$}; 
\node[circle,draw](C3) at (2,2){$C_3$}; 
\node[circle,draw](A) at (0,4){$A$};
\node[circle,draw](Y) at (4,4){$Y$};
\node[circle,draw](M) at (2,4){$M$}; 
% edges
\draw[->, thick]
(C1) edge (A)
(C1) edge (M)
(C1) edge (Y)
(C1) edge (C2)
(C2) edge [bend left=40] (A)  
(C2) edge (M) 
(C2) edge [bend right=30] (Y)
(C3) edge (M) 
(C3) edge (Y) 
(A) edge [bend left=30] (Y)
(A) edge (M)
(M) edge (Y);
\end{tikzpicture}
\end{document}
\end{lstlisting} & 
\begin{lstlisting}
library(dagitty)
dag2 <- dagitty('dag {
  A [exposure,pos="0.250,0.150"]
  C_3 [pos="1.025,0.600"]
  C_2 [pos="0.500,1.000"]
  C_1 [pos="0.500,0.600"]
  M [pos="0.900,0.150"]
  Y [outcome,pos="1.500,0.150"]
  A -> M
  A -> Y [pos="1.000,0.075"]
  C_1 -> C_2
  C_3 -> M
  C_3 -> Y
  C_2 -> A [pos="0.150,0.500"]
  C_2 -> M
  C_2 -> Y [pos="1.500,0.700"]
  C_1 -> A
  C_1 -> M
  C_1 -> Y
  M -> Y
}')

plot(dag2)
\end{lstlisting}
 & 
\begin{lstlisting}
library(ggdag)
dag2 <- dagify(
   # relationship for each node
    A ~ C1 + C2 +C3 ,
    C2 ~ C1,
    M ~ A + C1 + C2 + C3,
    Y ~ M + A + C1 + C2 + C3,
    # specify exposure and outcome 
    exposure = "A", outcome = "Y",
    # Location of each node
    coords = list(
      x = c(Y = 3, M = 2, A = 1, 
            C1 = 1, C2 = 1, C3 = 2),
      y = c(Y = 2, M = 2, A = 2,
            C1 = 1, C2 = 0, C3 = 1))
  ) 
dag2 %>% 
  ggplot(aes(x = x, y = y, 
    xend = xend, yend = yend)) +
  geom_dag_point(shape=1) + 
  #geom_dag_edges() + 
  geom_dag_edges_arc( 
    curvature = c(0, 0.35, 0, 0, 0,
        0, 0.35, 0, -0.35, 0, 0, 0)) +
  theme_void() +
  geom_dag_text(# subscripts
    color = "black",
    parse = TRUE, 
    label = c("A","M","Y",
        expression(C[1]),
        expression(C[2]),
        expression(C[3])) ) 
\end{lstlisting} & 
\begin{lstlisting}
library(dagR)
dag2 <- dag.init(
  outcome = NULL, 
  exposure = NULL,
  covs = c(1, 1, 1, 1), 
  arcs = # write in pairs
   c(1,0,  # C1 to A
     1,4,  # C1 to M
     1,-1, # C1 to Y
     1,2,  # C1 to C2
     2,0,  # C2 to A
     2,4,  # C2 to M
     2,-1, # C2 to Y
     3,4,  # C3 to M
     3,-1, # C3 to Y
     4,-1, # M to Y
     0, 4, # A to M
     0,-1  # A to Y
  ),
  # directed arcs, 0 per cov
  assocs = c(0, 0, 0, 0), 
  xgap = 0.04, 
  ygap = 0.05, 
  len = 0.05,
  symbols = c("A", "C_1", "C_2",
             "C_3","M", "Y"),
  noxy = T
)
plot_dag2 <- dag.draw(dag2, 
    noxy = T, legend = F)
\end{lstlisting} & 
\begin{lstlisting}
library(igraph)
dag2 <- make_graph(
  edges = c("A","M",#A to M
            "A","Y", #A to Y
            "M","Y", #M to Y
            "C1","A", #C1 to A
            "C1","M", #C1 to M
            "C1","Y", #C1 to Y
            "C2","A", #C2 to A
            "C2","M", #C2 to M
            "C2","Y", #C2 to Y
            "C3","M", #C3 to M
            "C3","Y"),#C3 to Y
  directed = TRUE)
plot(dag2, 
     layout = matrix(
         c(1, 2, 3, 1, 1, 2,
           2, 2, 2, 1, 0, 1),
         nrow = 6, ncol = 2
     ), #matrix of location
     edge.curved = c(
         0, 0.35, 0, 0, 0, 0, 
         0.4, 0, -0.4, 0, 0, 0), 
     vertex.size = 25, 
     vertex.label= c(
         "A", "M", "Y", 
         expression(C[1]), 
         expression(C[2]), 
         expression(C[3])), 
     vertex.color = "white", 
     edge.color = "black", 
     vertex.label.color = "black", 
     edge.arrow.size = 0.6 )
\end{lstlisting} \\  \bottomrule
    \end{tabular}
    \caption{Complex mediation DAG example for each software. $A$ denotes the exposure, $Y$ the outcome, $M$ the mediator, and ``C\_1''/$C_1$, ``C\_2''/$C_2$, ``C\_3''/$C_3$ the three covariates. The code used to generate each DAG is included below the relative plot.
    }
    \label{fig:dag2}
\end{figure}
\end{landscape}

To compare the \ap{five} software packages we create the same two DAGs using each program. In Figure \ref{fig:dag1}, we implement an identical simple confounding causal structure in each software and compare the output. In Figure \ref{fig:dag2}, we draw a more complex causal relationship including one mediator and three confounders. \cf{For Figures \ref{fig:dag1} and \ref{fig:dag2}, we include below the plots the code used to produce each DAG.} We create the DAGs using DAGitty (web version 3.0, CRAN version 0.3-1), ggdag (version 0.2.\ap{10}), dagR (version 1.2.1), \ap{and igraph (version 1.5.0.1)} with R-4.2.0 \cite{R}, and the Ti\textit{k}Z graphs in \LaTeX\ 2022 with Ti\textit{k}Z (version 3.1.10). In Table \ref{tab:comparetab}, we summarize the capabilities of each software based on their visual design, analysis capability\cf{,} and utility. \cf{We produce the graphs from the four R packages by saving the output as a PNG file, while the Ti\textit{k}Z graphs are produced in \LaTeX\ by compiling to a PDF. We note however that all five methods are compatible with R Markdown and Quarto. In Figure \ref{fig:dag1} and \ref{fig:dag2}, we attempt to maintain uniform styling across methods, with circles around nodes, black text and arrows, and nodes arranged chronologically from left to right. In Table \ref{tab:comparetab} we highlight design customization settings available for each software to further adapt the DAG to the user's preferences.}

\subsection{Ti\textit{k}Z}

Ti\textit{k}Z is a \LaTeX\ library for creating graphical figures, with extensive customization settings to generate a vast array of different images. It relies on PGF (portable graphics format), another \LaTeX\ language as its base layer \cite{tantau2023tikz}. Till Tantau developed and maintains the two libraries. Tantau did not design the library specifically for DAGs, and thus there is no built-in exposure, outcome, and covariates framework, analysis features, nor automatic placement of nodes. Despite these limitations, Ti\textit{k}Z  serves as one of the leading software to create DAGs, thanks to its flexibility which allows the user to easily specify the color, size, shape, and arrangement of nodes and edges. Additionally, as Ti\textit{k}Z is built into the \LaTeX\ environment, one can use mathematic\cf{al} notation to label variables with inline math mode (denoted by \$'s), including subscripts, \cf{G}reek letters, and any other mathematical symbols. There are many tutorials and user-posted question and answer boards online that explain the different possible customization settings. One can also refer to the Ti\textit{k}Z and PGF manual for the full documentation of all of the package's capabilities \cite{tantau2023tikz}. For the purpose of creating causal diagrams, we recommend beginning with DAG\cf{-}specific resources as Ti\textit{k}Z's extensive language can be overwhelming, especially for a beginner in \LaTeX. \cf{To code a DAG using Ti\textit{k}Z, we first create the nodes and specify each node's location, shape, color, and label as desired. We then list the edges and again can customize any stylistic preferences. The code is both intuitive and readable.}

In Figure \ref{fig:dag1} we write the labels using plain text, add circles around the nodes, and use the stealth arrow type. We display the variables chronologically and arrange the exposure, outcome, and confounder such that all edges can be shown using straight lines. In Figure \ref{fig:dag2} we use a similar design as in Figure \ref{fig:dag1}, but here write variable labels using inline math mode. This allows us to add subscripts to the confounder nodes, and label them as $C_1, C_2$, and $C_3$. Here we incorporate curved arrows, by specifying the angle of the desired curve. For those with less familiarity with \LaTeX, there may be a steeper learning curve to create figures with Ti\textit{k}Z. However, we believe that frequent users of \LaTeX\ will find it easy to incorporate Ti\textit{k}Z in their documents to produce DAGs. Looking at Figures \ref{fig:dag1} and \ref{fig:dag2}, it is obvious why Ti\textit{k}Z is a popular resource to draw DAGs. We easily \ap{can} create a clear and visually appealing DAG, with straightforward changes available to \cf{adapt} the style to a user's preferences. In Table \ref{tab:comparetab} we \cf{further see} that Ti\textit{k}Z's strength comes from its customizability and visual appeal, while its main limitation is from not being designed for a DAG\cf{-}specific framework and thus not having analysis capabilities.

\subsection{DAGitty}

DAGitty is a browser-based interface, downloadable tool, and R library for creating, editing, and analyzing DAGs. The website interface and downloadable tool are accessible via \href{http://www.dagitty.net/}{http://www.dagitty.net/} and the R package is available on CRAN \cite{textor2016robust}.  DAGitty's browser version provides a graphical user interface that allows users to draw and analyze causal diagrams. Its drag-and-click features make the tool very user-friendly and \cf{easy to learn}. The website allows the user to select and label nodes, connect nodes via directional edges, and identify the exposure, outcome, and covariates. After one creates a causal diagram, \cf{one} can explore the conditional independence, ancestor/descendant identification, and minimally sufficient adjustment sets \cf{\cite{van2014constructing, van2014constructing, van2015efficiently}}. The user can also copy and paste the model code into R after \cf{installing} and loading the DAGitty library. \cf{Similarly, t}he R library has \ap{the} functionality to obtain the conditional independencies, ancestor/descendant identification, and adjustment set lists directly in the statistical program. The DAGitty package in R additionally offers functions that can simulate data based on the specified DAG structure \cite{textor2016robust}. However, the simulation functionality is limited; the creators suggest \cf{employing} it only for validation purposes and that one \cf{use} other techniques or software for more complicated simulation studies \cite{textor2016robust, breitling2022dagrsim}. 
DAGitty 0.9a\cf{,} the oldest version of the software available was released in 2010 with its first announcement via a letter in \textit{Epidemiology} \cite{textor2011dagitty}. The most current version \cf{of the R package }available is \cf{3.1}, which was updated in \cf{2023} (as of \cf{August} 2023). DAGitty developers also maintain the browser-based website regularly \cite{dagitty_website}. 

In Figure \ref{fig:dag1} we see the simple confounding DAG with DAGitty's output \cf{from R}. Notably\ap{,} there are no circles around the nodes, the lines are very light and thin, and the arrows \cf{run} very close to the letters. Figure \ref{fig:dag2} shows the more complicated mediation DAG with DAGitty's graph shown \cf{second to the left}. We can see that DAGitty is not able to incorporate the subscripts on the nodes. It is able to plot the curved edges, however the placement and execution of the curves do not look as polished as in Ti\textit{k}Z. The top curved arrow from $A$ to $Y$ appears condensed due to space limitations imposed by \cf{the} RStudio plot output box and the R Markdown display region. If the curve had a larger radius from $A$ to $Y$ then the display region in R would cut off the top part of the curve. This region limitation is not a problem on the DAGitty website. 

To create both of these DAGs we employ the website to set up the initial placement and then copy and paste code from the website into RStudio, where we use the DAGitty R library. The DAGitty website is user-friendly and \cf{intuitive}.  Table \ref{tab:comparetab} shows that DAGitty does well as far as the analysis capability and utility but is less \cf{flexible} in its visual design. Overall, this is a \cf{suitable package} for users who want to produce a quick DAG and \cf{can accept compromises }on visual appearance.   

\subsection{ggdag}

The R package ggdag allows users to plot and analyze causal graphs \cite{barrett2018ggdag}. It is built on top of DAGitty to utilize DAGitty's powerful algorithms to analyze DAGs, while allowing users to employ ggplot and tidyverse to create professional, reproducible, and visually appealing DAGs \cite{barrett2020introduction}. It also enables the use of DAGitty objects in the context of tidyverse \cite{barrett2018ggdag, barrett2023ggdag}. The R functions are flexible in the sense that they allow users to code their DAG structure using DAGitty syntax or ggdag-specific syntax. This feature allows ggdag to have the same analytical \cf{capability} that DAGitty has\cf{,} including identification of conditional independencies, ancestor/descendant relationships, and minimally sufficient adjustment sets lists \cite{barrett2018ggdag}. Ggdag \cf{additionally} offers a visual display of adjustment sets via a colored graph \cite{barrett2022common, barrett2023ggdag}. \cf{Lastly, t}his package has a wrapper function that allows one to apply DAGitty's simulating data algorithm to the structural equation model \cite{ barrett2023ggdag}. Thus, ggdag and DAGitty are both able to simulate data, but under the same limitations. The initial release was in March of 2018 and the current version \cf{0.2.10} was updated in 2023 (as of \cf{August} 2023) \cite{barrett2023ggdag}.

Figure \ref{fig:dag1} displays the simple confounding DAG with ggdag's output shown in the \cf{third }panel. \ap{In ggdag, edges are specified with structural equation model notation in the \texttt{dagify()} function. The \texttt{ggplot()}  plotting function then renders the defined dagify object.}
Figure \ref{fig:dag2} illustrates the more complicated mediation DAG with ggdag's version displayed in \cf{center column}.  We see that ggdag is able to incorporate curved edges and subscripts on the node labels. The curved edges have a nice bold arc, for which it was easy to control the radius and directionality \ap{using the \texttt{geom\_dag\_edge\_arc()} add on}. The incorporation of both the subscript and the professional-looking curved edges make this graph visually appealing. The creation of this graph took twice the amount of time \cf{as} DAGitty\cf{, suggesting a longer} learning curve\cf{, however} this might be eased with more frequent use. Incorporating the placement of each node, the location of each curved edge, and the subscripts each took the authors multiple \cf{G}oogle searches for examples, vignettes, or user-posted question and answer boards \cite{barrett2020introduction, barrett2022introduction, barrett2022common, barrett2023ggdag}. The end product is very appealing but \cf{it} did require patience and time. 
\ap{For quicker use of ggdag, one might utilize DAGitty's web application to specify node and edge locations and then use the corresponding code and dagitty object in ggdag's plotting function.}

Table \ref{tab:comparetab} shows that ggdag performs relativity well as far as analysis capability, and visual appeal. Notably, ggdag can incorporate subscripts and curved arrows, customize node placement, and write Greek symbols to label nodes through unique \cf{U}nicode values in the ggdag label function \cite{rahlf2017data}. In the utility category, ggdag has many vignettes, is well documented, and has many resources available online \cite{barrett2023ggdag, barrett2020introduction, barrett2022introduction, barrett2022common}, however ggdag falls short compared to other software in \cf{ease-of-use} due to the relative time needed by the authors to create Figure \ref{fig:dag2}. Overall, \cf{ggdag} is \cf{a valuable tool} for users who want to create professional-looking DAGs in R. 

\subsection{dagR}
DagR \cf{is} an R package developed by Lutz P Breitling that was originally released in 2010 and most recently updated in 2022 (as of \ap{August} 2023) \cite{breitling2010dagr, breitling2022dagrsim}. The package allows users to plot DAGs, identify minimal sufficient adjustment sets, list ancestors of a given node, and simulate data from the specified causal diagram. DagR notably provides additional simulation capabilities compared to DAGitty (and ggdag) by allowing for a combination of binary and continuous variables within the same DAG \cite{breitling2022dagrsim}.  Unlike DAGitty and ggdag, dagR does not provide functions to identify conditional independencies, and cannot directly identify descendants \cite{breitling2010dagr}. 

While dagR excels in simulating data and provides some analysis functions, it has limitations compared to the other available software in terms of visualizations and user\cf{-}friendliness. The package is lacking \cf{in }customization settings for DAGs, as it does not allow for curved edges, circles around nodes, subscripts, nor math notation. In Figure \ref{fig:dag1}, we see the visual design is fairly similar to DAGitty, with smaller node labels and thin edges. The dagR default settings print a legend below the DAG, allowing the user to provide longer labels or descriptions for the nodes. However, we found the legend lines tend to overlay, reducing readability\cf{,} as seen in Figure \ref{fig:dag1}. In Figure \ref{fig:dag2}, \cf{we employ the automatic placement of nodes feature, as without curved arrows this achieves the clearest arrangement of all edges in the graph. Unfortunately, the DAG still has visual limitations, such as the overlayed arrow heads leading into variables $M$ and $Y$.} With more complex DAGs it would be challenging to generate a readable graph without curved lines. 

Additionally\cf{,} we note that \ap{d}agR, despite the simple visual results, is one of the hardest to \cf{utilize}. Compared to DAGitty and ggdag, there are few online resources available with example code and \cf{use }descriptions. We also find the code itself to be the least intuitive. For example, \cf{all }edges are specified \cf{together} in a single vector using pairs of numbers, where each number refers to a different node. With a larger number of variables\cf{,} it becomes difficult to track which number corresponds to each node, making the system \cf{cumbersome}. In Table \ref{tab:comparetab} we highlight dagR's strength in analysis and particularly data simulation, but limitations in visual design and utility. The package \cf{lags} in terms of the readability of its graphs and code. Lastly, we note that dagR developers provide a function to translate dagR objects to DAGitty \cite{breitling2022dagrsim}. Thus, if a user desires a more visually appealing graph, but has already written their causal structure using dagR (perhaps to simulate data) they can convert the object to DAGitty, and use DAGitty or ggdag to plot.

\subsection{igraph}

\ap{ Lastly, we include igraph, an open-source network analysis tool that emphasizes efficiency and portability \cite{csardi2006igraph}. Gábor Csárdi and Tamás Nepusz began the development of igraph in 2006 \cite{csardi2006igraph}, however many collaborators have since contributed to its growth. The most recent R update of igraph as of August 2023 is version 1.5.0.1, released in July 2023 \cite{igraph_cran}. The tool can be utilized in R, Python, Mathematica, and C/C++ but its core is written in C \cite{csardi2006igraph}. Here, we focus on igraph's implementation in R. This package excels at auto-placement of nodes and edges utilizing its vast array of graph layout algorithms, making it a valuable tool for visualizing complex and large DAGs \cite{csardi2006igraph, igraphInterface}.  
While igraph provides many functions to calculate various structural properties of graphs and conduct network analysis,  it has limited causal analysis capabilities. It can identify ancestors and descendants (i.e. \texttt{subcomponent()}), but does not offer functions to directly compute conditional independencies and adjustment sets. \cite{igraphInterface}.}

\ap{The igraph notation to specify arrows is similar to dagR, where edges are listed in a vector with every pair denoting the origin and destination node of the edge \cite{csardi2006igraph, igraphInterface}. However, unlike in dagR, one can name nodes with characters, making the code more readable. While automatic arrangement of the nodes is the default, the user can specify the x and y coordinates for each node via a layout matrix. In Figure \ref{fig:dag1}, we see that the igraph DAG allows for black text surrounded by a circle for each node with black lines connecting each node. Since the default visualization uses navy text, orange circles, and gray arrows, we use several add-on features to achieve the desired styling (i.e. \texttt{edge.color}, \texttt{vertex.size} and \texttt{vertex.color} are used in the plotting function). In Figure \ref{fig:dag2} we again specify the color, size, label, and location of all nodes and edges. To draw curved lines, we use the \texttt{edge.curved} specification. To handle subscript notation we use \texttt{vertex.label} and the \texttt{expression()} feature; one could easily specify greek letters here using Unicode. A full list of plotting controls is conveniently located in the R Vignette \cite{igraphInterface}. The DAGs produced are clean, professional, and highly customizable, although fine-tuning all of the plotting features can be time-consuming. In summary, igraph excels in its flexibility for the visual design of a DAG, however it has limited analysis functions to answer causal inference questions, and less intuitive code compared to Ti\textit{k}Z, DAGitty, and ggdag. }

\section{Discussion}

There are several ways to create DAGs using open-source software, each with different strengths and weaknesses. By focusing on two DAG structures (seen in Figure\ap{s} \ref{fig:dag1} and \ref{fig:dag2}) we are able to compare the software and highlight key \cf{features}. Our findings are summarized \cf{in} Table \ref{tab:comparetab} by displaying the software's visual design, analysis capability\cf{,} and utility.  In this review\cf{,} we \cf{choose} to focus on only \ap{5} software: Ti\textit{k}Z, DAGitty, ggdag\ap{, dagR and igraph,} and primarily restrict our scope to directed acyclic graphs. 

\cf{The graphs we create in Figures \ref{fig:dag1} and \ref{fig:dag2} attempt to employ circular nodes, all-black coloring, and chronological arrangement but we note that there is some debate on best stylistic practices for DAGs. While some recommend that variables be arranged such that arrows flow in a single direction (e.g. left to right or top to bottom)\cite{tennant2021use, greenland1999dagdesc}, others arrange nodes by causal proximity \cite{miao2018identifying}. Additionally, some choose to limit circular nodes to latent or unobserved variables, and use square nodes or no shape for observed variables \cite{tennant2021use, silva2006principled}. In Table \ref{tab:comparetab}, we highlight which software would be adept at these design changes.}  To create an acyclic directed mixed graph (ADMG) \cite{richardson2003markov} all the software highlighted except for dagR \cf{can} display bidirected edges as needed. \cf{Meanwhile, Ti\textit{k}Z-SWIG, a library using Ti\textit{k}Z in \LaTeX, is the leading software to draw Single World Intervention Graphs (SWIGs \cite{richardson2013single})
\cite{ Richardson_2021}.} \cf{More} work is needed to provide a synopsis of software available to generate these other types of causal graphs and evaluate performance.

We recommend that one choose \cf{a package} to draw DAGs \cf{appropriate to one's} needs. Should the reader want to quickly produce an informal graph, we suggest using DAGitty, as the online platform allows one to generate a figure without writing any code. For DAG interpretations such as identifying minimally sufficient adjustment sets, ancestor/descendant relationships, and conditional independencies, DAGitty offers the widest range of functionality. When simulating data from a DAG, dagR provides the most flexibility. \ap{For visualizing large and complex DAGs we recommend using igraph}. \ap{Lastly, for formal publication quality graphs we recommend Ti\textit{k}Z when the manuscript is being written in \LaTeX\ and the rendered output is a PDF, and ggdag when the final result will be coded in R utilizing an R Markdown or Quarto compiler.} 

While the discussed software offer a wide range of capabilities, we are optimistic that drawing DAGs will become even easier as new tools arise and existing \cf{software} improve. \ap{We write this paper using the listed versions of each of the software; the discussed packages may have future updates that address some of the shortcomings we identify.} \cf{We encourage developers and users to continue contributing to the open-source DAG software community and look forward to future developments.} We anticipate that soon the option will exist to take a photo of a hand-drawn graph and convert it to code \cf{to render} the DAG in digital form. Such software already exists for mathematical formula\cf{s}, matrices, and chemical diagrams \cite{costa2021mathpix, mathpixweb}, and it is the \cf{logical next} step for DAG drawing tools.

\section{Acknowledgments}

Thank you to Caleb Miles, Linda Valeri, and Daniel Malinsky for their invaluable and constructive feedback. We would also like to extend our appreciation to the Columbia University Causal Inference Learning Group for creating a supportive environment and \cf{facilitating }the informative discussions. \cf{We express our gratitude to the developers, blog and tutorial authors, and thread posters and commenters for providing such a rich basis of tools and knowledge for using software to work with DAGs. We would lastly like to thank our reviewers for their instructive and insightful commentary and feedback.}

\bibliographystyle{vancouver}
\bibliography{sources}

\newpage 
\lstset{style=myCustomRStyle}

%
%
%


\begin{landscape}
\begin{table}

    \centering
    \scalebox{0.78}{
    \begin{tabular}[c]{ 
    >{\raggedright\arraybackslash}p{0.2\linewidth}|
    >{\raggedright\arraybackslash}p{0.2\linewidth}|
    >{\raggedright\arraybackslash}p{0.2\linewidth}|
    >{\raggedright\arraybackslash}p{0.2\linewidth}|
    >{\raggedright\arraybackslash}p{0.2\linewidth}|
    >{\raggedright\arraybackslash}p{0.2\linewidth}} \toprule 
         & \textbf{Ti\textit{k}Z} & \textbf{DAGitty} & \textbf{ggdag} & \textbf{dagR} & \textbf{igraph} \\ \hline 
         \rowcolor{Gray}
    \multicolumn{6}{l}{\normalsize \textbf{Visual Design}} \\ \hline 
        
        \textbf{Curved arrows} & 
        \texttt{[bend} \textit{direction}\texttt{=}\textit{angle}\texttt{]} & 
        \texttt{[pos=``}\textit{\#}\texttt{,}\textit{\#}\texttt{'']} after edge specification & 
        \texttt{+ geom\_dag\_edge\_arc()} & 
        - & 
        \texttt{edge.curved=c()} in \texttt{plot()}\\ 
        \hline
        
        \textbf{Subscripts} & 
        Underscore in inline math mode for node label (\texttt{\$\_\$}) &  
        - &  
        \texttt{expression()} in \quad\quad\quad\quad\texttt{+ geom\_dag\_text()} & 
        - & 
        \texttt{vertex.label=c()} in \texttt{plot()}\\ 
        \hline
        
        \textbf{Math notation}  &  
        Inline math mode for node label (\texttt{\$\$})  &  
        - & 
        \texttt{+ geom\_dag\_text()} using Unicode & 
        - & 
        \texttt{vertex.label=c()} in \texttt{plot()}\\ 
        \hline
        
        \textbf{Auto-generated placements} & 
        - &  
        This is default option in the plotting function &  
        This is default option in the plotting function &  
        This is default option in the plotting function &  
        This is default option in the plotting function\\ 
        \hline
        
        \textbf{Manual placements} & 
        \texttt{at (}\textit{x}\texttt{,}\textit{y}\texttt{)} after node command OR placement relative to other nodes \texttt{[}\textit{direction}\texttt{=of }\textit{other node}]& 
        \texttt{[pos=``x,y"]} after node specification OR give a list of x and y pairs in the \texttt{coordinates()}&  
        \texttt{coords = list(x=c(), y=c())} in the \texttt{dagify()} function & 
        \texttt{dag\$x=c()} for x coordinates and \texttt{dag\$y=c()} for y coordinates & 
        \texttt{layout = matrix()} in \texttt{plot()} where each row is the corresponding (x,y) coordinate\\ 
        \hline
        
        \textbf{Default visual settings} & 
        Black text and arrows, no circles   &  
        In R: black text and arrows, no circles; on website: gray circles and black text and gray arrows & 
        Black shaded circles with white text, black arrows & 
        Black text and arrows, no circles & 
        Navy text, orange circles, gray arrows \\  
        \hline

        \textbf{Design customizations} & 
        Easy to add circles/shapes and change color, size, and weight of nodes and edges   &  
        In R: cannot add shapes around nodes, cannot customize color; on website: custom color, can remove circles  & 
        Easy to customize shape, color, size, and location of nodes and edges & 
        No circles available, cannot customize color & 
        Easy to customize/remove shapes, can change color, size, and location of nodes and edges \\ \hline   
        
        \rowcolor{Gray}
    \multicolumn{6}{l}{\normalsize \textbf{Analysis Capability} } \\ \hline 
        \textbf{Exposure, outcome  framework} & 
        - &  
        Specify the \texttt{[exposure]} or \texttt{[outcome]} after each node & 
        Specify the \texttt{exposure} and \texttt{outcome} in the \texttt{dagify()} function & 
        Outcome and exposure automatically included as node \#\texttt{-1,0} respectively & 
        -\\ 
        \hline
        
        \textbf{Ancestor / descendant identification}   &  
        - &  
       \multicolumn{2}{l|}{\texttt{ancestors()}, \texttt{descendants()}}      
       & 
        \texttt{dag.ancestors()} & 
        \texttt{subcomponent()}; with \texttt{mode="in"} or \texttt{mode="out"} \\ 
        \hline
        
        \textbf{Conditional independencies identification}   & 
        -  & 
        \multicolumn{2}{l|}{ \texttt{impliedConditionalIndependencies()} }  
         &  
        - & 
        -\\ 
        \hline
        
        \textbf{Adjustment sets calculation}   &  
        - & 
        \texttt{adjustmentSets()} &  
        \texttt{ggdag\_adjustment\_set()}& 
        \texttt{dag.search()} & 
        -\\ 
        \hline
        
        \textbf{Data Simulation} &  
        - & 
        \texttt{simulateSEM()} &  
        \texttt{simulate\_data()}& 
        \texttt{data.sim()}& 
        -\\ \hline  

        \rowcolor{Gray}
    \multicolumn{6}{l}{\normalsize \textbf{Utility} }\\ \hline  
        \textbf{Resources / online help} & 
        \cite{tantau2023tikz, Grantham_2022} &   
        \cite{textor2016robust, textor2011dagitty, dagitty_website, van2014constructing,van2015efficiently} &   
        \cite{barrett2023ggdag, barrett2020introduction, barrett2022introduction, barrett2022common} & 
         \cite{breitling2010dagr, breitling2022dagrsim} & 
        \cite{csardi2006igraph, igraph_cran, igraphInterface}\\ 
        \hline
        
        \textbf{Learning curve} & 
        Straightforward code, relatively easy to customize & 
        Straightforward to use website, easy to copy and paste code & 
        Straightforward to use defaults, longer learning curve to customize & 
        Longer learning curve required to specify nodes and edges, less intuitive & 
        Longer learning curve required to specify edges, straightforward to plot\\ 
        \hline
        
        
        \textbf{Base software} & 
        \LaTeX  & 
        website / R &  
        R & 
        R & 
        C/C++, R, Python \\ \bottomrule
    \end{tabular} }
    \caption{Summary of the characteristics and capabilities for each reviewed software. Where possible, the relative code/functions are mentioned. Note that '-' is used to denote when a feature is not available. 
    }
    \label{tab:comparetab}
\end{table}
\end{landscape}

\end{document}